\documentclass[twocolumn]{webofc}
\usepackage[varg]{txfonts}   % Web of Conferences font
\usepackage{tikz}
\usepackage{hyperref}
\usepackage{wrapfig}
\usepackage{url}
\usepackage{subcaption}
\hypersetup{colorlinks=true,citecolor=blue,urlcolor=blue,linkcolor=blue}
\begin{document}
\vspace{-0cm}
\title{Quantification of Oxygen and Carbon in Calcium Targets for Reliable Ca$(p,p\alpha)$ Measurements}
\author{\firstname{Junki} \lastname{Tanaka}\inst{1,2}\fnsep\thanks{\email{junki@rcnp.osaka-u.ac.jp}} \and
        \firstname{Riku} \lastname{Matsumura}\inst{3,4}\fnsep
            \and  
        \firstname{Taichi} \lastname{Miyagawa}\inst{1,2}\\
             for the ONOKORO collaboration
        % etc.
}
\institute{Research Center for Nuclear Physics (RCNP), The University of Osaka, Osaka 567-0047, Japan
\and
Department of Physics, The University of Osaka, Osaka 560-0043, Japan
\and
           Graduate School of Science and Engineering,
Saitama University, Saitama 338-8570, Japan
\and
           RIKEN Nishina Center for Accelerator-Based Science, Saitama 351-0198, Japan
          }

\abstract{
Reliable extraction of Ca$(p,p\alpha)$ cross sections requires accurate correction
for oxygen and carbon impurities in calcium targets.
In this work, the relative amounts of these light elements in
$^{40,42,44,48}$Ca targets are determined using 65-MeV proton elastic scattering,
where the Ca/Mylar yield ratios provide a direct measure of the corresponding
O and C atomic ratios.
These experimentally determined ratios are then applied to the 392-MeV
$(p,p\alpha)$ spectra to subtract the O and C contributions in a fully
data-driven manner.
The method does not rely on assumptions about absolute contamination levels
or reaction-model calculations, and enables a consistent and reliable
determination of Ca$(p,p\alpha)$ yields across the calcium isotopic chain.
}
\maketitle
The $\alpha$-knockout method has become a powerful probe of cluster correlations
in the nuclear surface region~\cite{Tanaka2021}, where the reduced nuclear density
is expected to enhance the emergence of cluster degrees of freedom.
This approach can be extended to a wide range of nuclei and reaction channels,
including $t$, $^3$He, and $d$ knockout, with the aim of establishing a unified
description of cluster degrees of freedom in medium- to heavy-mass systems
\cite{Uesaka2024,Kubota2025}.

A practical experimental challenge in Ca$(p,p\alpha)$ measurements is the presence of small oxygen and carbon impurities introduced during target preparation.
Although their absolute concentrations are low, the corresponding O$(p,p\alpha)$ and C$(p,p\alpha)$ cross sections are significantly larger than those of Ca, reflecting the lighter masses and more weakly bound cluster structures of these nuclei.
In addition, their $\alpha$-separation energies ($S_\alpha$) are close to those of the Ca isotopes.
As a consequence, O/C$(p,p\alpha)$ events appear in the same $S_\alpha$ region as the Ca$(p,p\alpha)$ signals and cannot be distinguished by kinematic selection alone.
A quantitative evaluation of these contributions is therefore essential for reliable isotopic comparisons.

Table~\ref{tab:targets} summarizes the thicknesses, isotope enrichment, and $\alpha$-separation energies of the Ca targets used in the present study. 

\begin{table}[h!]
    \centering
    \begin{tabular}{cccc}
        \hline
        Target & Thickness & Enrichment & $S_\alpha$\\
        Isotope & (mg/cm$^2$) & (\%) & (MeV)\\\hline
        $^{40\rm(nat)}$Ca & 11.8 (1.1) & 96.9 & 7.039\\
        $^{42}$Ca         & 10.0 (1.0) & 91.2 & 6.257\\
        $^{44}$Ca         &  9.7 (1.0) & 88.5 & 8.854\\
        $^{48}$Ca         & 10.0 (1.0) & 97.8 & 13.977\\\hline
    \end{tabular}
    \caption{
    Thicknesses, isotope enrichment, and $\alpha$-separation energies
    of the Ca targets used in this work.
    }\vspace{-0.6cm}
    \label{tab:targets}
\end{table}

In the present work, the relative amounts of $^{16}$O and $^{12}$C in the Ca
isotopic targets are determined using 65-MeV proton elastic-scattering
measurements.
These measurements were performed under identical experimental conditions
for the Ca isotopes and the Mylar reference target, allowing a direct comparison
of elastic-scattering yields.

\begin{figure}
  \centering
  \includegraphics[width=0.99\linewidth]{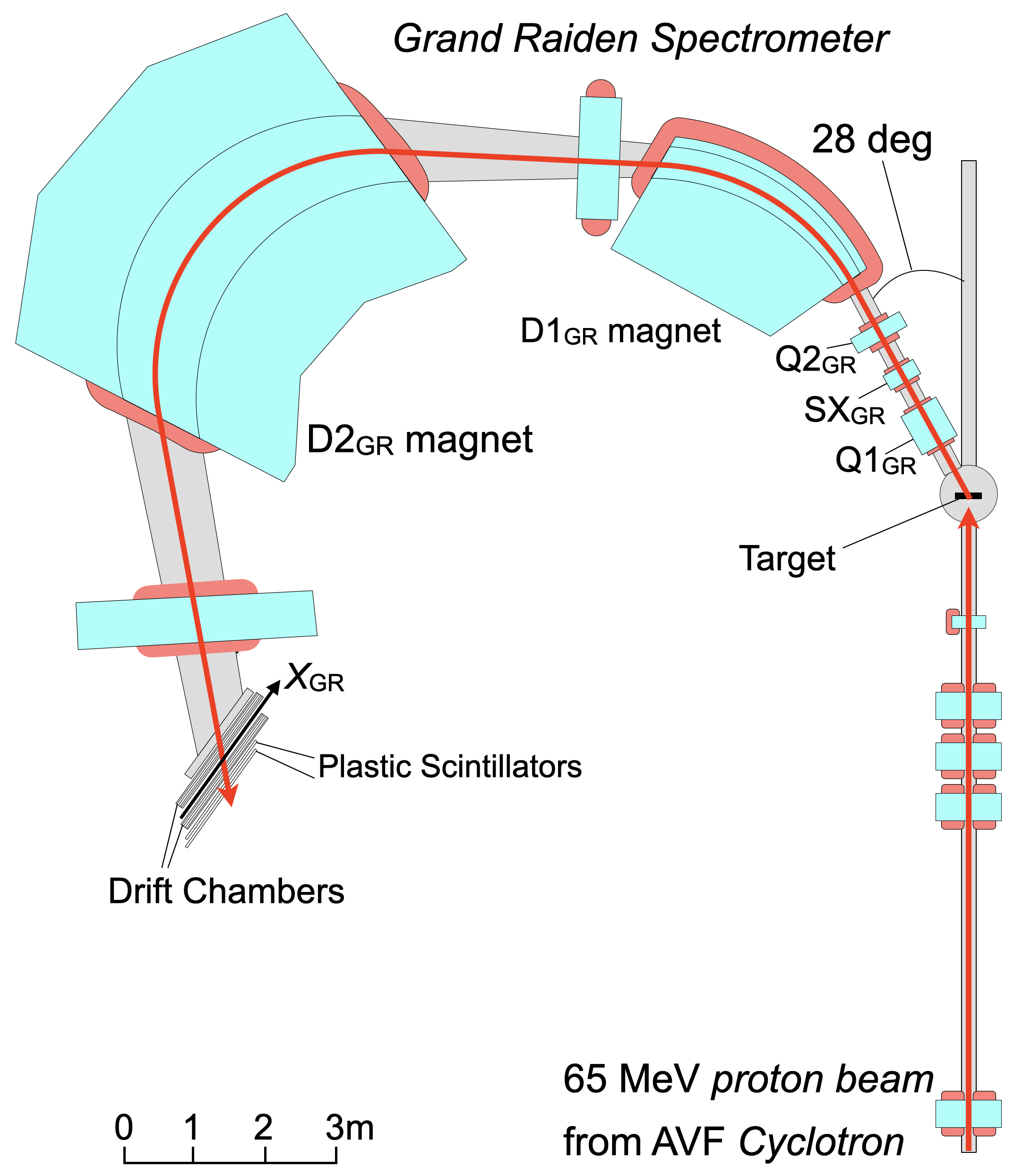}
  \caption{
The Grand Raiden spectrometer was positioned at a laboratory angle of
$\theta_{\rm lab}=28^\circ$ with respect to the beam axis.
Elastically scattered protons were transported to the focal plane and measured
using drift chambers for position determination and plastic scintillators for
energy-loss and timing measurements.
Identical spectrometer settings were employed for the Ca and Mylar targets to
ensure identical kinematic and acceptance conditions.
  }
\end{figure}

The essential point of this approach is that the ratio of the
$^{16}$O$(p,p)$ elastic-scattering yields per unit luminosity between the Ca and
Mylar targets depends solely on the relative number of oxygen atoms and is
independent of the reaction mechanism and absolute cross sections.
The same procedure is applied to $^{12}$C.
These experimentally determined ratios are directly used in the subsequent
392-MeV $(p,p\alpha)$ analysis to quantify and subtract the O/C contributions
in the Ca spectra, without relying on assumptions about absolute contamination
levels.
This strategy provides a consistent and robust correction procedure for
extracting reliable Ca$(p,p\alpha)$ yields across the calcium isotopic chain.
In the present work, the relative amounts of $^{16}$O and $^{12}$C in the Ca
isotopic targets are determined using 65-MeV proton elastic-scattering
measurements performed under identical experimental conditions for the Ca and
Mylar targets.
Under these identical kinematic and instrumental conditions, the ratio of the
$^{16}$O$(p,p)$ elastic-scattering yields per unit luminosity between the two
targets depends only on the relative number of oxygen atoms and is therefore
independent of the reaction mechanism and absolute cross sections.
These experimentally determined ratios are directly applied to the 392-MeV
$(p,p\alpha)$ analysis to quantify the O/C contributions in the Ca spectra,
without relying on absolute contamination levels.
This strategy ensures a consistent and robust correction procedure for
extracting reliable Ca$(p,p\alpha)$ yields across the Ca isotopic chain.
Elastic-scattering measurements were carried out at RCNP using a 65-MeV proton
beam accelerated by the AVF cyclotron.
The Grand Raiden spectrometer~\cite{Fujiwara1999} was set at a laboratory angle of
$\theta_{\rm lab} = 28^\circ$, and its magnetic fields were optimized for the
detection of elastically scattered protons.
Data were collected for $^{40\rm(nat)}$Ca, $^{42}$Ca, $^{44}$Ca, $^{48}$Ca, and
Mylar targets under identical experimental conditions, enabling a direct and
consistent comparison of the O$(p,p)$ and C$(p,p)$ yields.
The energy acceptance of approximately 10\% allowed elastically scattered
protons from all target nuclei to be detected within a single spectrometer
setting.
To ensure comparable energy resolution between the Ca and Mylar measurements,
the Mylar foil thickness was chosen such that the resulting energy straggling
was essentially identical to that of the Ca targets.
As a result, the energy resolution of the 65-MeV elastic-scattering spectra was
nearly identical for all targets, with a typical value of
$\mathrm{FWHM} \approx 50$~keV, primarily determined by the energy spread of the
incident beam.

\begin{figure}[h!]
  \vspace{-0.2cm}
  \centering
  \includegraphics[width=\linewidth]{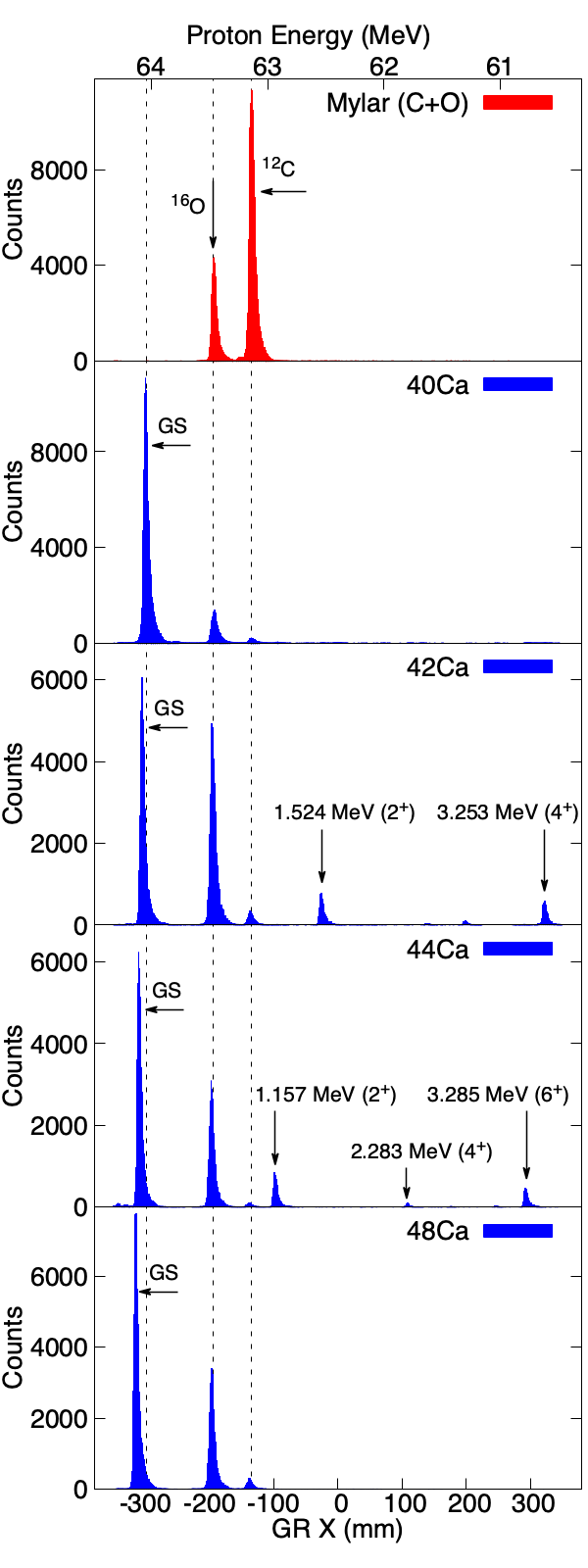}
  \vspace{-0.4cm}
  \caption{
  Grand Raiden focal-plane $X$ position distributions corresponding to the
  energy spectra of protons from 65-MeV elastic scattering on the Mylar,
  $^{\rm nat}$Ca, $^{42}$Ca, $^{44}$Ca, and $^{48}$Ca targets.
  }\vspace{-1.0cm}
  \label{fig:elastic_spectrum}
\end{figure}

Figure~\ref{fig:elastic_spectrum} shows the resulting spectra.
The lower horizontal axis represents the $X$ position at the focal plane of the
Grand Raiden spectrometer, which is converted to the proton energy on the upper
axis.
In the Mylar spectrum (top), prominent peaks originating from elastic scattering
on ${}^{16}$O and ${}^{12}$C are observed.
The ${}^{40}$Ca spectrum (second) exhibits a dominant elastic ground-state (GS)
peak, accompanied by smaller contributions from ${}^{16}$O and ${}^{12}$C.
The spectra for ${}^{42}$Ca (third) and ${}^{44}$Ca (fourth) show, in addition to
the ground state, several inelastic peaks that are assigned to known excited
states based on their energies.
For ${}^{48}$Ca (bottom), the first $2^+$ state lies outside the present energy
acceptance, and only the ground-state peak is observed.
A systematic shift of the Ca elastic peaks reflects the change in isotope mass
and confirms both the isotope selection and the high isotopic purity of the
targets.
In all spectra, the ${}^{16}$O and ${}^{12}$C peaks appear at identical energies,
as indicated by the dotted reference lines.

The oxygen content in each target is determined from the
${}^{16}$O$(p,p)$ yields $Y_{{}^{16}{\rm O}(p,p)}$, which are related to the elastic
differential cross section $(d\sigma/d\Omega)_{\rm el}$ by
\begin{align}
Y^{\rm (Ca)}_{{}^{16}{\rm O}(p,p)} &=
\left(\frac{d\sigma}{d\Omega}\right)_{\rm el}
Q^{\rm (Ca)} N^{\rm (Ca)}_{{}^{16}{\rm O}}\,\Delta\Omega, \\
Y^{\rm (My)}_{{}^{16}{\rm O}(p,p)} &=
\left(\frac{d\sigma}{d\Omega}\right)_{\rm el}
Q^{\rm (My)} N^{\rm (My)}_{{}^{16}{\rm O}}\,\Delta\Omega .
\end{align}
Here, $Q$ is the accumulated beam charge measured during the experiment,
$N$ denotes the number of ${}^{16}$O atoms in each target, and
$\Delta\Omega$ is the solid angle of the spectrometer.

Because the Ca and Mylar targets were measured under identical kinematic and
instrumental conditions, taking the yield ratio eliminates the dependence on
the absolute cross section and the solid angle:
\begin{equation}
\frac{N^{\rm (Ca)}_{{}^{16}{\rm O}}}{N^{\rm (My)}_{{}^{16}{\rm O}}} =
\frac{Y^{\rm (Ca)}_{{}^{16}{\rm O}(p,p)}}{Y^{\rm (My)}_{{}^{16}{\rm O}(p,p)}}
\cdot
\frac{Q^{\rm (My)}}{Q^{\rm (Ca)}} .
\end{equation}

This ratio depends solely on the relative number of oxygen atoms in the two
targets.
This reaction-independent property provides the basis for applying the 65-MeV
normalization factors directly to the 392-MeV $(p,p\alpha)$ analysis.
Table~\ref{tab:O16_summary} summarizes the measured ${}^{16}$O$(p,p)$ yields, the
effective beam charges, and the resulting oxygen ratios.
The same procedure is applied to ${}^{12}$C, whose contamination level is
determined using an identical ratio formulation.

\vspace{-0.2cm}
\begin{table*}[h]
  \centering
  \caption{Summary of $^{16}$O/$^{12}$C$(p,p)$ yields, beam charges, and the ratios to the Mylar target.}
  \label{tab:O16_summary}\vspace{-0.2cm}
  \begin{tabular}{lcccccc}
    \hline
    Target & $Y_{^{16}\mathrm{O}(p,p)}$ & $Y_{^{12}\mathrm{C}(p,p)}$  & $Q$ (nC) & $N^{\rm (Target)}_{{}^{16}{\rm O}}/N^{\rm (My)}_{{}^{16}{\rm O}}$ & $N^{\rm (Target)}_{{}^{12}{\rm C}}/N^{\rm (My)}_{{}^{12}{\rm C}}$ \\
    \hline
    Mylar     & 547026 & 1286616 &7148 & 1.0000 (0.0026)& 1.0000 (0.0021) \\
%    natC      &   5615 & 77  & 3352.1  & 3.4  & 0.0219 & 0.0003 \\
    $^{40}$Ca & 211571 & 30235 & 29277 & 0.0944 (0.0003)& 0.0057 (0.0001) \\
    $^{42}$Ca & 645655 & 44770 & 18626 & 0.4530 (0.0011)& 0.0134 (0.0001) \\
    $^{44}$Ca & 357845 & 13657 & 16477 & 0.2838 (0.0008)& 0.0046 (0.0001) \\
    $^{48}$Ca & 410519 & 38687 & 16103 & 0.3331 (0.0009)& 0.0133 (0.0001) \\
    \hline
  \end{tabular}
\end{table*}

With the number ratios of ${}^{16}$O and ${}^{12}$C relative to the Mylar target
established at 65~MeV, their contributions to the 392-MeV $(p,p\alpha)$ spectra
are evaluated.
A 65-MeV proton beam from the AVF cyclotron was subsequently accelerated to
392~MeV by the Ring Cyclotron.
Using a double-arm spectrometer operated under identical settings,
$(p,p\alpha)$ reactions were measured for the Ca isotopes, Mylar, and an
additional ${}^{\rm nat}$C target.
Protons were detected with the Grand Raiden spectrometer at $45^\circ$, and
$\alpha$ particles with the Large Acceptance Spectrometer (LAS) at $58^\circ$,
corresponding to quasi-free kinematic conditions.

For each target, $\alpha$-separation-energy $S_\alpha$ spectra were constructed
using the missing-mass method and normalized by the accumulated beam charge.
Details of the analysis procedure are described in Ref.~\cite{Tanaka2021}.

The yields of the ${}^{16}$O$(p,p\alpha)$ $S_\alpha$ spectra for the Ca and Mylar
targets are expressed as
\begin{equation}
\frac{Y^{(\rm Ca)}_{^{16}{\rm O}(p,p\alpha)}}{Q^{(\rm Ca)}} =
\left(
\frac{d^{3}\sigma}{d\Omega_p\, d\Omega_\alpha\, dS_\alpha}
\right)_{^{16}{\rm O}(p,p\alpha)}
N^{(\rm Ca)}_{^{16}{\rm O}}\,
\Delta\Omega_p\, \Delta\Omega_\alpha\, \Delta S_\alpha ,
\label{eq:4}
\end{equation}
\begin{equation}
\frac{Y^{(\rm My)}_{^{16}{\rm O}(p,p\alpha)}}{Q^{(\rm My)}} =
\left(
\frac{d^{3}\sigma}{d\Omega_p\, d\Omega_\alpha\, dS_\alpha}
\right)_{^{16}{\rm O}(p,p\alpha)}
N^{(\rm My)}_{^{16}{\rm O}}\,
\Delta\Omega_p\, \Delta\Omega_\alpha\, \Delta S_\alpha ,
\label{eq:5}
\end{equation}
where $\Delta\Omega_p$ and $\Delta\Omega_\alpha$ denote the solid-angle
acceptances of the proton and $\alpha$ spectrometers, respectively, and
$\Delta S_\alpha$ is the bin width of the $S_\alpha$ spectrum.

The reaction kinematics is uniquely determined once the proton scattering angle
$\Omega_p$, the $\alpha$ emission angle $\Omega_\alpha$, and one additional
kinematic variable are specified.
In the present analysis, the $\alpha$-separation energy $S_\alpha$, reconstructed
using the missing-mass method, is chosen as this third variable.
With this choice, each $(\Omega_p,\Omega_\alpha,S_\alpha)$ bin corresponds to a
unique reaction kinematics, and the yield in that bin is proportional to the
triple-differential cross section.

The triple-differential cross section
$(d^3\sigma/d\Omega_p\,d\Omega_\alpha\,dS_\alpha)_{^{16}{\rm O}(p,p\alpha)}$
depends only on the ${}^{16}$O$(p,p\alpha)$ reaction mechanism and kinematics.
Because the Ca and Mylar measurements were performed with the same incident beam
energy, spectrometer settings, and angular acceptances, the reaction conditions
for ${}^{16}$O are identical in both targets.
Consequently, the triple-differential cross section is common to
Eqs.~\ref{eq:4} and \ref{eq:5}.

Equation~\ref{eq:4} represents the ${}^{16}$O$(p,p\alpha)$ yield arising from the
${}^{16}$O impurities in the Ca target.
Because this contribution appears at the same $S_\alpha$ values as the
Ca$(p,p\alpha)$ reaction, it overlaps with the Ca spectrum and cannot be isolated
from the Ca-target data alone.

Equation~\ref{eq:5} gives the ${}^{16}$O$(p,p\alpha)$ yield obtained from the
Mylar target.
In the Mylar data, the ${}^{16}$O and ${}^{12}$C contributions can be separated,
since the ${}^{12}$C$(p,p\alpha)$ spectrum is independently determined.
Subtracting this known ${}^{12}$C contribution therefore provides a reference
${}^{16}$O$(p,p\alpha)$ spectrum free of Ca-related components.

Taking the ratio of Eqs.~\ref{eq:4} and \ref{eq:5} cancels the spectrometer
acceptances $\Delta\Omega_p\Delta\Omega_\alpha$, the $S_\alpha$ bin width
$\Delta S_\alpha$, and the absolute
${}^{16}$O$(p,p\alpha)$ triple-differential cross section.
The resulting relation depends only on the relative number of ${}^{16}$O atoms
in the two targets:
\begin{equation}
\frac{Y^{(\rm ^{40}Ca)}_{^{16}{\rm O}(p,p\alpha)}}{Q^{(\rm ^{40}Ca)}} =
\left(
\frac{N_{^{16}{\rm O}}^{({\rm ^{40}Ca})}}
     {N_{^{16}{\rm O}}^{({\rm My})}}
\right)
\frac{Y^{(\rm My)}_{^{16}{\rm O}(p,p\alpha)}}{Q^{(\rm My)}} ,
\end{equation}
\begin{equation}
\frac{Y^{(\rm ^{44}Ca)}_{^{16}{\rm O}(p,p\alpha)}}{Q^{(\rm ^{44}Ca)}} =
\left(
\frac{N_{^{16}{\rm O}}^{({\rm ^{44}Ca})}}
     {N_{^{16}{\rm O}}^{({\rm My})}}
\right)
\frac{Y^{(\rm My)}_{^{16}{\rm O}(p,p\alpha)}}{Q^{(\rm My)}} .
\end{equation}

\begin{figure}[h!]
  \centering
  \begin{tikzpicture}
    \node[anchor=south west, inner sep=0] (img)
      {\includegraphics[width=0.99\linewidth]{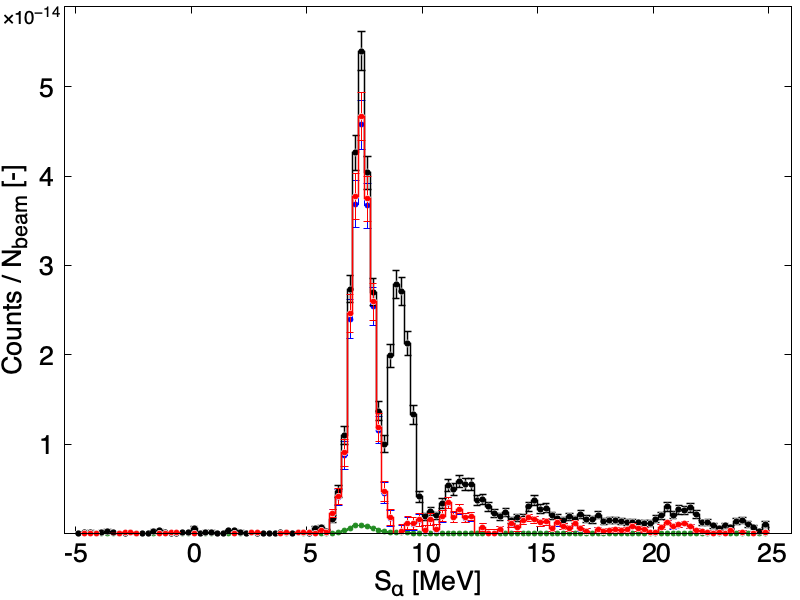}};
    \node[
      anchor=north west,
      font=\bfseries\Large,
      text=red!50,
      xshift=45mm,
      yshift=-5mm
    ] at (img.north west) {Preliminary};
  \end{tikzpicture}
  \caption{
$(p,p\alpha)$ $S_\alpha$ spectrum for the ${}^{44}$Ca target.
The black curve shows the measured spectrum.
The blue and green curves indicate the estimated
${}^{16}$O$(p,p\alpha)$ and ${}^{12}$C$(p,p\alpha)$ contributions, respectively,
scaled from the Mylar reference data.
The red curve represents their sum.
  }\vspace{-0.6cm}
  \label{fig:ppalpha44}
\end{figure}

\begin{figure}[h!]
  \centering
  \begin{tikzpicture}
    \node[anchor=south west, inner sep=0] (img)
      {\includegraphics[width=0.99\linewidth]{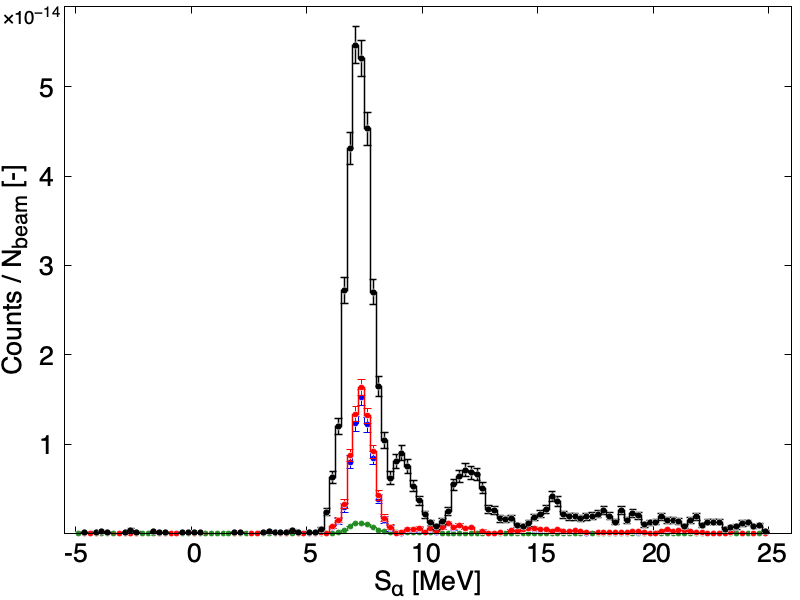}};
    \node[
      anchor=north west,
      font=\bfseries\Large,
      text=red!50,
      xshift=45mm,
      yshift=-5mm
    ] at (img.north west) {Preliminary};
  \end{tikzpicture}
  \caption{
$(p,p\alpha)$ $S_\alpha$ spectrum for the ${}^{40}$Ca target.
The black curve shows the measured spectrum.
The blue and green curves indicate the estimated
${}^{16}$O$(p,p\alpha)$ and ${}^{12}$C$(p,p\alpha)$ contributions, respectively,
scaled from the Mylar reference data.
The red curve represents their sum.
  }\vspace{-0.6cm}
  \label{fig:ppalpha40}
\end{figure}

Figures~\ref{fig:ppalpha40} and \ref{fig:ppalpha44} show the measured
$(p,p\alpha)$ $S_\alpha$ spectra for ${}^{40}$Ca and ${}^{44}$Ca, respectively,
together with the estimated oxygen and carbon contributions obtained using the
procedure described above.
The yield is normalized per incident proton, where the measured beam charge $Q$
is converted to the number of incident particles $N$ via $Q=eN$.

As shown in Fig.~\ref{fig:ppalpha44}, for the ${}^{44}$Ca target the
O$(p,p\alpha)$ and C$(p,p\alpha)$ components are distributed around
$S_\alpha \simeq 7$~MeV, while the
${}^{44}{\rm Ca}(p,p\alpha){}^{40}{\rm Ar}$ peak is located near
$S_\alpha \simeq 9$~MeV.
The separation between these components is about 2~MeV in $S_\alpha$.
Under this condition, the summed O/C contribution scaled from the Mylar reference
is overlaid on the Ca-target spectrum without additional normalization.
The difference between the O/C yields extracted from the Ca-target data and those
estimated from the Mylar reference provides an estimate of the associated
systematic uncertainty. Benchmarking this procedure against independent data sets yields a
reproducibility of 7\%.

In contrast, Fig.~\ref{fig:ppalpha40} shows that for ${}^{40}$Ca the
${}^{40}{\rm Ca}(p,p\alpha){}^{36}{\rm Ar}$ peak is located in the same
$S_\alpha$ region as the oxygen and carbon contributions, around
$S_\alpha \simeq 7$~MeV.
In this region, the estimated O/C contribution amounts to approximately 25\% of
the total yield.
Propagating the 7\% uncertainty of the O/C estimation therefore results in a
systematic uncertainty of
$0.25 \times 0.07 \simeq 0.02$, corresponding to a 2\% uncertainty in the
extracted ${}^{40}$Ca$(p,p\alpha)$ yield.
This contribution is smaller than other dominant systematic
uncertainty, such as those arising from the target thickness.

Overall, the consistency between the Ca-target data and the Mylar-scaled O/C
contributions, quantified by the reproducibility of the extracted yields,
supports the validity of the data-driven procedure used to quantify oxygen and
carbon impurities. Because the method relies exclusively on experimentally determined yield ratios, it does not require assumptions on absolute contamination levels or
reaction models. The resulting uncertainty satisfies the precision requirements of the present analysis and is consistently applied across the calcium isotopic chain.

%\vspace{8pt}
\noindent\textbf{Acknowledgments} \;
The authors thank the accelerator group at RCNP for their technical support. 
This work was supported by JSPS KAKENHI (JP21H04975) and the JSPS A3 Foresight Program ``Nuclear Physics in the 21st Century.'' Additional support was provided by the NRF TOPTIER Korea--Japan Joint Research Program (RS-2024-00436392), IBS (IBS-R031-D1) in Korea, and the National Key R\&D Program of China (2023YFE0101500).
\vspace{-30pt}
\end{document}